\documentclass[aps,prl,twocolumn,superscriptaddress,floatfix]{revtex4}
\usepackage{graphicx}
\usepackage{epsfig}
\usepackage{amsmath}
\usepackage{amssymb}
\usepackage{natbib}
\usepackage{times}

\begin{document}

\title{Negative Hydration Expansion in ZrW$_{\mathbf 2}$O$_{\mathbf 8}$:\\ Microscopic Mechanism, Spaghetti Dynamics, and Negative Thermal Expansion}

\author{Mia Baise}
\affiliation{Department of Chemistry, University of Oxford, Inorganic Chemistry Laboratory, South Parks Road, Oxford OX1 3QR, U.K.}
\affiliation{Department of Chemistry, University College London, Gower Street, London WC1E 6BT, United Kingdom}

\author{Phillip M. Maffettone}
\affiliation{Department of Chemistry, University of Oxford, Inorganic Chemistry Laboratory, South Parks Road, Oxford OX1 3QR, U.K.}

\author{Fabien Trousselet}
\affiliation{Chimie ParisTech, PSL Research University, CNRS, Institut de Recherche de Chimie Paris, 75005 Paris, France}

\author{Nicholas P. Funnell}
\affiliation{Department of Chemistry, University of Oxford, Inorganic Chemistry Laboratory, South Parks Road, Oxford OX1 3QR, U.K.}
\affiliation{ISIS Neutron and Muon Facility, Rutherford Appleton Laboratory, Didcot OX11 0QX, U.K.}

\author{Fran{\c c}ois-Xavier Coudert}
\affiliation{Chimie ParisTech, PSL Research University, CNRS, Institut de Recherche de Chimie Paris, 75005 Paris, France}

\author{Andrew L. Goodwin$^\ast$}
\affiliation{Department of Chemistry, University of Oxford, Inorganic Chemistry Laboratory, South Parks Road, Oxford OX1 3QR, U.K.}

\date{\today}

\begin{abstract}
We use a combination of X-ray diffraction, total scattering and quantum mechanical calculations to determine the mechanism responsible for hydration-driven contraction in ZrW$_2$O$_8$. Inclusion of H$_2$O molecules within the ZrW$_2$O$_8$ network drives the concerted formation of new W--O bonds to give one-dimensional (--W--O--)$_n$ strings. The topology of the ZrW$_2$O$_8$ network is such that there is no unique choice for the string trajectories: the same local changes in coordination can propagate with a large number of different periodicities. Consequently, ZrW$_2$O$_8$ is heavily disordered, with each configuration of strings forming a dense aperiodic `spaghetti'. This new connectivity contracts the unit cell \emph{via} large shifts in the Zr and W atom positions. Fluctuations of the undistorted parent structure towards this spaghetti phase emerge as the key NTE phonon modes in ZrW$_2$O$_8$ itself. The large relative density of NTE phonon modes in ZrW$_2$O$_8$ actually reflect the degeneracy of volume-contracting spaghetti excitations, itself a function of the particular topology of this remarkable material.
\end{abstract}

\maketitle

Zirconium tungstate, ZrW$_2$O$_8$, is an important material because it harbours a number of useful and counterintuitive mechanical responses to external stimuli. First, on heating, its molar volume \emph{decreases} such that the material is 2\% denser at 1050\,K than at 0.3\,K \cite{Martinek_1968,Evans_1996,Mary_1996,Evans_1999}. This negative thermal expansion (NTE) effect is about as strong as the usual positive thermal expansion (PTE) of conventional materials, and so can be exploited in athermal composites \cite{Evans_1999b,Barerra_2005,Lind_2012}. Second, under hydrostatic pressure, ZrW$_2$O$_8$ amorphises unusually easily, transforming into a disordered phase that is denser by more than 20\% \cite{Perottoni_1998,Varga_2005,Keen_2007}. This switching between crystalline and amorphous states is related to the mechanism exploited for data storage in phase-change chalcogenides \cite{Hegedus_2008}. And, third, hydration of ZrW$_2$O$_8$ gives the monohydrate ZrW$_2$O$_8\cdot$H$_2$O, which has a molar volume $\sim$10\% smaller than ZrW$_2$O$_8$ itself \cite{Duan_1999,Banek_2010}. In this sense the material---which has no appreciable pore network---is nonetheless a kind of `inverse sponge' that shrinks instead of swells when it takes up water. We call this effect ``negative hydration expansion'' (NHE). NHE is known to occur in anisotropic `breathing' frameworks such as MIL-53 \cite{Serre_2002,Neimark_2010}, but is rare in cubic systems \cite{Bousquet_2013,Krause_2016}; pore sizes are typically 5--20\,\AA\ in all cases. The phenomenon is largely unexplored---and particularly so in ZrW$_2$O$_8$---yet has potential applications in counteracting hydration-driven swelling of construction materials (\emph{e.g.}\ concretes and clays).

While the mechanism responsible for NHE in ZrW$_2$O$_8$ remains unclear \cite{Duan_1999,Banek_2010}, both NTE and pressure-induced amorphisation (PIA) have been studied extensively and are understood to arise from a peculiar phonon spectrum. The material supports an extensive family of low-energy modes with large and negative Gr{\"u}neisen parameters \cite{Ramirez_1998,Mittal_1999,Wang_2000,Gava_2012,Gupta_2013,Rimmer_2015} (\emph{i.e.} their frequencies soften with pressure)---the thermodynamic requirement for NTE in cubic systems \cite{Barerra_2005}. Under hydrostatic pressure the energies of these modes fall to zero \cite{Perottoni_1998,Mittal_2001}. Because the modes are distributed throughout the Brillouin zone (BZ), the resulting structural transition is associated with a family of modulation periodicities, and hence amorphisation \cite{Keen_2007}.

Despite the many experimental and computational studies of ZrW$_2$O$_8$, the microscopic origin of its phonon anomalies remains contentious. It is the sheer number of NTE vibrational modes that is difficult to explain: each one appears to involve a remarkably different combination of polyhedral translations, rotations, and distortions \cite{Gupta_2013}. Early indications of the importance of `rigid unit modes' (RUMs) \cite{Pryde_1996,Tao_2003,Tucker_2005} have increasingly lost experimental \cite{Cao_2002,Bridges_2014} and computational \cite{Gupta_2013,Rimmer_2015} support, with consensus only emerging regarding the role of polyhedral distortions and the `skipping rope' or `tension' mechanism proposed in Ref.~\citenum{Mary_1996}.

Intriguingly, hydration switches off NTE in ZrW$_2$O$_8$: ZrW$_2$O$_8\cdot$H$_2$O is a conventional PTE material \cite{Banek_2010,Ahmad_2014}. One implication is that the mechanism responsible for NHE saturates the same set of displacements from which NTE arises. Consequently we sought to determine the structure of ZrW$_2$O$_8\cdot$H$_2$O, establish a mechanism for NHE, and explore the relationship between NHE and NTE in ZrW$_2$O$_8$ itself.

In this Letter, we show that hydration of ZrW$_2$O$_8$ proceeds via concerted formation of one-dimensional strings of new W--O bonds. The topology of the ZrW$_2$O$_8$ network is such that there is no unique choice for the trajectory of these strings: the same local changes in coordination can propagate with a macroscopically large number of different periodicities. Each configuration of strings forms a dense `spaghetti' that describes an additional connectivity responsible for contracting the unit cell \emph{via} surprisingly large shifts in the Zr and W atom positions. So ZrW$_2$O$_8\cdot$H$_2$O is heavily disordered, albeit in a highly correlated fashion. Remarkably, fluctuations of the anhydrous parent towards this spaghetti phase emerge as the key NTE phonon modes in ZrW$_2$O$_8$. Consequently the density and distribution of NTE modes throughout reciprocal space actually reflect the degeneracy of volume-contracting spaghetti excitations, itself a function of network topology.

\begin{figure}
\begin{center}
\includegraphics{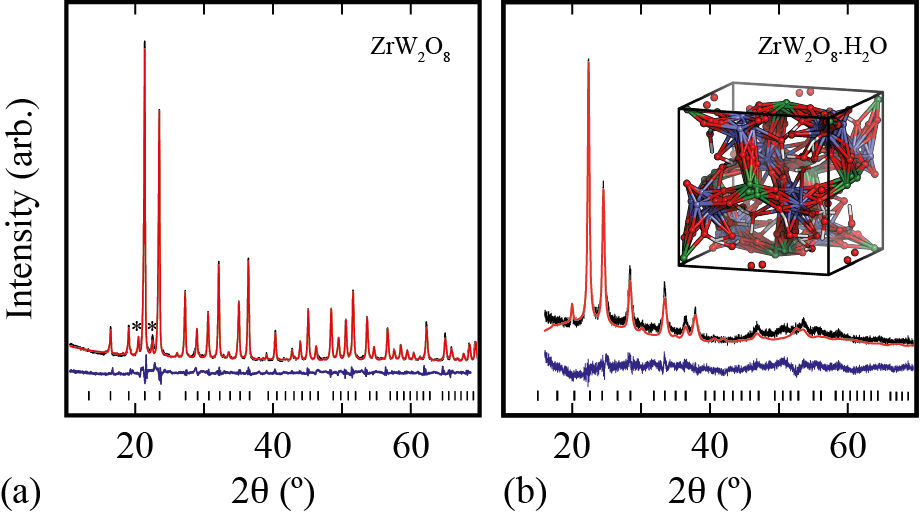}
\end{center}
\caption{\label{fig1} X-ray diffraction patterns ($\lambda=1.54056$\,\AA) of (a) anhydrous and (b) hydrated ZrW$_2$O$_8$: data in black, fits in red, difference (data$-$fit) in blue, and reflection positions as tick marks. The fit in (a) is from a Rietveld refinement using the coordinates of Ref.~\citenum{Mary_1996}; $^\ast$ = minor impurity. The fit in (b) used our DFT coordinates, with $Pa\bar3$ symmetry applied (inset: the corresponding structural model represents the configurational average of a disordered spaghetti).}\end{figure}

The starting point for our study was to prepare and characterise polycrystalline samples of both ZrW$_2$O$_8$ and its monohydrate, following the hydrothermal methodologies of Ref.~\citenum{Banek_2011}. The corresponding X-ray powder diffraction patterns [Fig.~\ref{fig1}] reflect all hydration-driven structural responses noted in earlier studies \cite{Duan_1999,Banek_2010}: the crystal symmetry increases on hydration from $P2_13$ to $Pa\bar3$, the lattice parameter $a$ decreases from 9.153(3) to 8.876(4)\,\AA, and there is a marked increase strain-driven peak broadening. We found hydration/dehydration cycles ($\sim$36\,h) to be reversible in all respects, including the appearance and disappearance of strain effects. Rietveld refinement of the parent ZrW$_2$O$_8$ diffraction pattern using the published structure \cite{Mary_1996} gave an excellent fit, but we encountered the same difficulty noted in \cite{Duan_1999} of obtaining a structural model for the monohydrate: the scattering density at the Zr and W sites indicated severe positional disorder and the O atom positions were impossible to determine.

\begin{figure}[b]
\begin{center}
\includegraphics{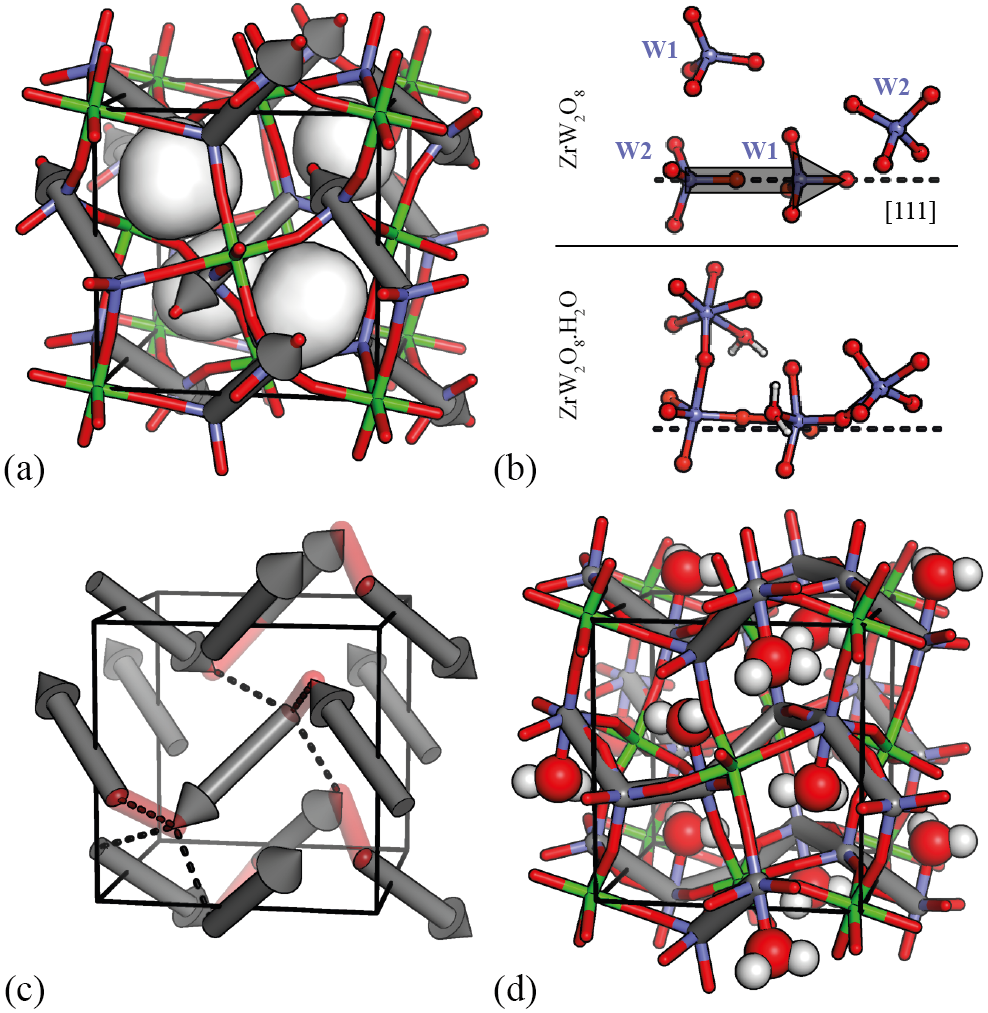}
\end{center}
\caption{\label{fig2} (a) The crystal structure of ZrW$_2$O$_8$: Zr in green, W in blue, O in red, and the four cavities of each unit cell in surface representation. (b) DFT geometries of the key structural fragment before (top) and after (bottom) hydration. Note the formation of additional W--O--W bonds. W2$\ldots$W1 pairs are indicated by a grey arrow; the location of these pairs within the structure is shown in (a). (c) On hydration, neighbouring W2$\ldots$W1 pairs connect to form W--O--W-- chains. Note the symmetry-equivalent choices for these new connections (dashed lines). The red cylinders denote the connections made in the DFT ($P2_12_12_1$) structure. (d) The final $P2_12_12_1$ structure, with W--O--W strings shown in grey and H$_2$O molecules shown in space-filling representation.}
\end{figure}

Given the difficulty of directly interpretating the X-ray diffraction pattern of ZrW$_2$O$_8\cdot$H$_2$O, we used quantum mechanical calculations to determine an energetically-sensible candidate structure for this phase. The structure of ZrW$_2$O$_8$ contains a single symmetry-related set of cavities, with each cavity of the correct size to host one H$_2$O molecule [Fig.~\ref{fig2}(a)]. There is one cavity per formula unit, and four formula units per unit cell. The cavities are positioned on the $4a$ Wyckoff sites and coincide with three-fold rotation axes of the $P2_13$ space group. Clearly this point symmetry is incompatible with that of the H$_2$O molecule. So there is a choice of three possible orientations for each of the four water molecules to be placed in a single unit cell. This situation contrasts that of NH$_3$ binding in ZrW$_2$O$_8\cdot$NH$_3$, for which NH$_3$ orientations are ordered and $P2_13$ symmetry is maintained \cite{Cao_2016}---the molecular point symmetry of NH$_3$ ($3m$) being a supergroup of the $4a$ site symmetry ($3$) \cite{nh3note}. Instead the arrangement of H$_2$O molecules that results in the least severe symmetry lowering gives $P2_12_12_1$ symmetry. We will come to show that ZrW$_2$O$_8\cdot$H$_2$O is in fact heavily disordered, with $Pa\bar3$ symmetry emerging as a configurational average over states related to this $P2_12_12_1$ arrangement we now consider in detail.

Starting with this highest-symmetry H$_2$O arrangement, we carried out density functional theory (DFT) relaxations at two different levels of theory (LDA and PBESOL0; see \cite{SI}). Energy minimisation was sensitive to the precise starting positions of the additional H$_2$O molecules within the framework cavities, yet we were able to identify a series of closely-related low-energy structures with molar volumes smaller by 6--8\% than that of ZrW$_2$O$_8$. The corresponding enthalpies of water adsorption were in the range 60--90\,kJ\,mol$^{-1}$ (PBESOL0 functional), indicating strong physisorption \cite{Bolis_2013}. Hence our calculations are entirely consistent with a NHE effect of the same magnitude as that observed experimentally. 

The structure of our lowest-energy $P2_12_12_1$ model for ZrW$_2$O$_8\cdot$H$_2$O reveals how hydration brings about a contraction of the unit cell. In this model, each H$_2$O molecule is chemically bonded to one of the two crystallographically-distinct W atoms [W1 in Fig.~\ref{fig2}(b)] along a direction that is no longer parallel to a triad axis of the original unit cell. This new coordination induces a series of coupled rotations and translations of both types of W-centred polyhedra that collectively results in a set of new W--O--W connections. These connections draw the framework in on itself: the terminal W1--O and W2--O bonds of the original structure are forced within bonding distance of nearby W2 and W1 atoms, respectively. Not only does this connectivity reduce the average W$\ldots$W separation---and in turn the unit-cell volume---but it results in substantial displacement of both Zr and W atoms away from their original sites. Indeed, provided we can account for the $Pa\bar3$ crystal symmetry, these displacements are fully consistent with experiment: Fig.~\ref{fig1}(b) shows the excellent match between the experimental diffraction pattern and that derived from our DFT coordinates. Moreover, our model accounts for the NTE/PTE switch on hydration: zone-centre phonon calculations show a dramatic reduction in the number of NTE modes, with the corresponding coefficient of thermal expansion now positive \cite{SI}.

Given the existence of this low-energy $P2_12_12_1$ structure, why would the experimental system retain cubic symmetry? We suggest the key is the existence of a diverging number of different ways of propagating the local changes in coordination identified by our DFT calculations. This degeneracy, we claim, leads to a strongly disordered state. We proceed to describe this state and in turn test our claims using experimental X-ray pair distribution function (PDF) measurements.

Consider the new W--O--W bonds formed during hydration [Fig.~\ref{fig2}(b)]. While there is only one W1 atom accessible to each terminal W2--O bond, there are three equally-spaced W2 atoms which any terminal W1--O bond might approach. These three choices of W2 atom are related by the same triad axis that is broken by the symmetry of an H$_2$O molecule. So, in selecting one of three possible states, the particular orientation of each H$_2$O molecule effectively determines the trajectory of a corresponding W1--O$\rightarrow$W2 linkage [Fig.~\ref{fig2}(c)]. These new W--O--W bonds form one-dimensional strings that in our DFT structure reflect the same $P2_12_12_1$ crystal symmetry of the H$_2$O molecule arrangements [Fig.~\ref{fig2}(d)]. But more disordered H$_2$O arrangements are associated with equally disordered arrangements of W--O--W strings (we will call collections of strings a `spaghetti', noting the relationship to the `loop' models \cite{Jaubert_2011} of \emph{e.g.}\ Coulomb phases \cite{Henley_2010,Jaubert_2012}). Such states are not random. Note that in the DFT structure each W1--O bond approaches only one W2 atom, and each W2 atom is approached by only one W1--O bond. Preserving these local constraints enforces correlations between the new W--O--W bonds formed, and in turn between the orientations of successive H$_2$O molecules. It can be shown \cite{SI} that the number of configurations satisfying these local rules grows exponentially with system size to give a macroscopic number of viable hydration states; the corresponding configurational entropy is $S_{\rm config}\gtrsim R\ln(9/8)$. Hence there is an entropic driving force for ZrW$_2$O$_8\cdot$H$_2$O to adopt spaghetti states, with the bulk crystal symmetry reflecting the (cubic) configurational average. There is a robust analogy between this model and icelike states on the pyrochlore lattice, which also preserve cubic symmetry \cite{Moessner_2006,Henley_2010,Overy_2016}.



In order to test the relevance of this spaghetti model to the NHE mechanism of ZrW$_2$O$_8\cdot$H$_2$O, we proceeded to measure X-ray total scattering data for both ZrW$_2$O$_8$ and its hydrate. Such data are directly sensitive to the local structure changes induced by hydration. We then developed a supercell model of a disordered spaghetti state, informed by the DFT structure we have already described, from which an X-ray-weighted PDF might be calculated and compared against experiment.

\begin{figure}[b]
\begin{center}
\includegraphics{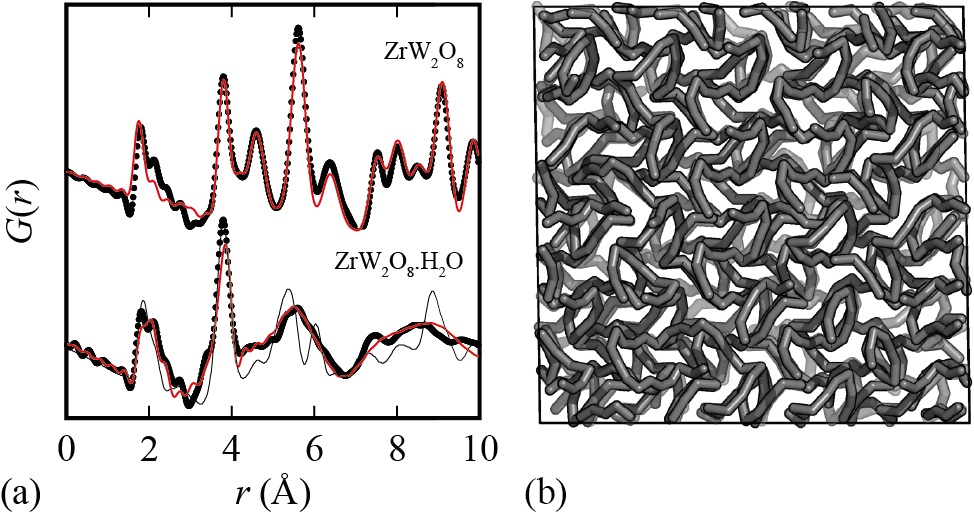}
\end{center}
\caption{\label{fig3} (a) X-ray PDFs $G(r)$ of anhydrous (top) and hydrated (bottom) ZrW$_2$O$_8$: data as points and fits as lines. For ZrW$_2$O$_8$ a satisfactory fit is obtained using the model of Ref.~\citenum{Mary_1996} (red lines). For the hydrate, the DFT $P2_12_12_1$ model fits at low $r$ but cannot reproduce the broad oscillations at $r>4$\,\AA\ (thin black line). Instead, the PDF calculated from our disordered spaghetti model (red line) captures all features. (b) The relaxed {\sc gulp} model for a ZrW$_2$O$_8\cdot$H$_2$O spaghetti configuration as described in the text: only W--O--W linkages are shown, using grey cylinders [\emph{cf}.~Fig.~\ref{fig2}(d)].}\end{figure}

Our total scattering measurements were performed using a PANalytical Empyrean X-ray diffractometer fitted with an Ag anode ($Q_{\rm max}=20$\,\AA$^{-1}$); the corresponding PDFs \cite{Keen_2001}, normalised using {\sc gudrunx} \cite{Soper_2011}, are shown in Fig.~\ref{fig3}(a). We were able to model the PDF of ZrW$_2$O$_8$ in {\sc pdfgui} \cite{Farrow_2007} using the crystal structure of Ref.~\citenum{Mary_1996}. By contrast, the $P2_12_12_1$ model for ZrW$_2$O$_8\cdot$H$_2$O is capable of fitting convincingly only the very lowest-$r$ region of the corresponding PDF, and cannot account for the positional disorder evident at distances $r\gtrsim4$\,\AA.

Because it was not computationally feasible to explore a disordered spaghetti configuration at the DFT level, we used harmonic lattice-dynamical methods, informed by the bonding geometries observed in our $P2_12_12_1$ DFT structure, to generate a $5\times5\times5$ approximant \cite{Thygesen_2017} [Fig.~\ref{fig3}(b)]. First we determined an appropriate disordered spaghetti of self-avoiding W--O--W pathways using a custom Monte Carlo algorithm, guided by the connectivity rules of Fig.~\ref{fig2}(c). This spaghetti was used to assign both connectivity and H$_2$O orientations within a supercell of the $P2_13$ ZrW$_2$O$_8$ structure. Atomic positions and unit cell dimensions were then relaxed using the {\sc gulp} code \cite{Gale_1997,SI}, with harmonic potentials constructed from the equilibrium bond distances and angles of our earlier DFT calculation. The cell volume decreased during relaxation, reflecting activation of the NHE mechanism as described above. We consider this {\sc gulp}-relaxed configuration to be representative of the degree of structural disorder implied by the spaghetti model. The connectivity of the relaxed structure is shown in Fig.~\ref{fig3}(b), and the PDF calculated from the full model compared against experiment in Fig.~\ref{fig3}(a). The quality of this fit-to-data shows that the displacive disorder introduced by spaghetti W--O--W connectivity can indeed account for the broad oscillations in the experimental PDF.

A necessary consequence of the macroscopic configurational entropy of spaghetti formation on the ZrW$_2$O$_8$ lattice is that the corresponding hydration-driven distortions are dense throughout the BZ. Hence the structure of ZrW$_2$O$_8\cdot$H$_2$O reflects a superposition of spaghetti states with arbitrarily large periodicities. Anisotropy vanishes in this limit such that the corresponding global symmetry is cubic, with an average structure given by the projection of all possible distortions onto a single unit cell [Fig.~\ref{fig1}(b)]. The large displacements involved account for the substantial crystal strain observed experimentally, and also for the difficulty of interpreting even high-quality neutron diffraction data \cite{Duan_1999}. In turn, proton migration and topological defects in spaghetti state formation account for the introduction of inversion symmetry. Consequently we conclude that NHE is driven by a degenerate family of `spaghetti'-like lattice distortions modulated with periodicities distributed continuously throughout the BZ.

\begin{figure}
\begin{center}
\includegraphics{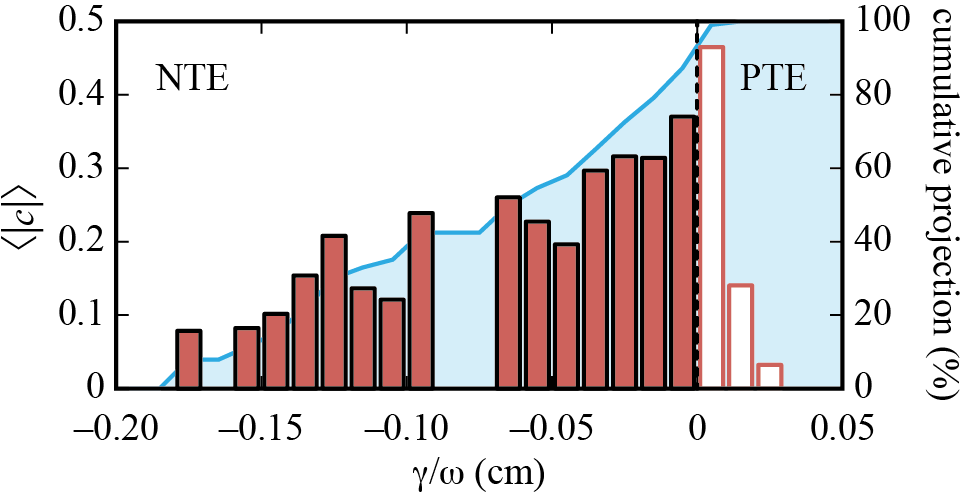}
\end{center}
\caption{\label{fig4} Hydration-driven framework displacements in ZrW$_2$O$_8\cdot$H$_2$O project strongly onto NTE phonon eigenvectors in ZrW$_2$O$_8$. The average magnitude of projection coefficients $c$ (vertical bars) is shown as a histogram of $\gamma/\omega$; nearly 90\% of the cumulative projection (blue shaded region) is accounted for by NTE modes.}\end{figure}

Having established this mechanism, we sought to determine whether the same distortions that drive NHE are related to the phonons responsible for NTE in ZrW$_2$O$_8$. Published data \cite{Rimmer_2015} allow access to the frequencies $\omega(\mathbf k,\nu)$, eigenvectors $\mathbf e(\mathbf k,\nu)$, and Gr{\"u}neisen parameters
\begin{equation}
\gamma(\mathbf k,\nu)=-\frac{V}{\omega(\mathbf k,\nu)}\frac{\partial}{\partial V}\omega(\mathbf k,\nu)
\end{equation}
for each phonon branch $\nu$ and wave-vector $\mathbf k$. NTE modes are those for which $\gamma<0$. By comparing the atomic coordinates of each ZrW$_2$O$_8$ framework atom in the anhydrous and hydrated DFT structures, we formed a normalised, mass-weighted displacement vector $\mathbf u$ that describes the volume-reducing framework deformations induced by hydration. The projections $c(\mathbf k,\nu)=\mathbf u\cdot\mathbf e(\mathbf k,\nu)$ partition these deformations amongst the normal modes; for the ($\mathbf k=\boldsymbol0$) DFT structure the $c(\mathbf k,\nu)$ necessarily vanish for $\mathbf k\neq\boldsymbol0$. Remarkably, we find that $88\%$ of the displacements described by $\mathbf u$ can be accounted for in terms of two of the lowest-energy $\mathbf k=\boldsymbol0$ NTE modes. The same result is found for zone-centre phonons generated by our quantum mechanical calculations \cite{SI}. Extending our analysis to wave-vectors $\mathbf k\neq\boldsymbol0$ requires access to supercell configurations. So, using the same {\sc gulp} approach described above, we generated a family of ten different $3\times3\times3$ spaghetti configurations. For each ZrW$_2$O$_8\cdot$H$_2$O configuration we extracted the corresponding framework displacement vectors $\mathbf u$ and projected these onto the phonon eigenvectors $\mathbf e(\mathbf k,\nu)$ of ZrW$_2$O$_8$ for $\mathbf k\in\langle\frac{n_1}{3},\frac{n_2}{3},\frac{n_3}{3}\rangle^\ast$, $n_i\in(0,\pm1)$. We find that the projections onto the very lowest energy modes (mostly acoustic) account for $\sim$45\% of the hydration-driven displacement, and on including all modes with negative Gr{\"u}neisen parameters this total projection rises to 87\% of the total framework response [Fig.~\ref{fig4}] \cite{SI}.

So the most important NTE phonon branches are strongly implicated in the NHE response of ZrW$_2$O$_8$, but the question remains of whether the spaghetti displacements invoked for NHE are necessarily involved in the NTE dynamics. The $c(\mathbf k,\nu)$ play a crucial role in answering this question: whereas previous studies have focussed either on individual collective modes \cite{Gupta_2013} or on the correlated motion of isolated structural fragments \cite{Bridges_2014}, the $c(\mathbf k,\nu)$ allow us to reconstruct combination NTE modes with intuitive real-space realisations. For example, the $P2_12_12_1$ DFT configuration of Fig.~\ref{fig2}(b) gives a framework displacement vector $\mathbf u$ that---as we have said---projects almost entirely onto just two zone-centre modes $\nu_1,\nu_2$ (note $\omega_1,\omega_2=44.1$\,cm$^{-1}$ and $\gamma_1,\gamma_2=-4.14$). The combination mode $c(0,\nu_1)\nu_1+c(0,\nu_2)\nu_2$ then describes one specific collective vibration of ZrW$_2$O$_8$ contributing to NTE. Crucially, the associated displacement vector $\mathbf e^\prime=c(0,\nu_1)\mathbf e(0,\nu_1)+c(0,\nu_2)\mathbf e(0,\nu_2)$ does indeed involve fluctuations towards the very same spaghetti states assumed by ZrW$_2$O$_8\cdot$H$_2$O; an animation of this combination mode is provided as SI \cite{SI}. The same result also holds for the more complex family of NTE modes associated with the $3\times3\times3$ spaghetti configurations described above \cite{SI}.

The significance of this result is that the key real-space mechanism for NTE in ZrW$_2$O$_8$ is the correlated translation, rotation and distortion of ZrO$_6$/WO$_4$ polyhedra towards the volume-reducing spaghetti states adopted in ZrW$_2$O$_8\cdot$H$_2$O. The configurational degeneracy of these states---a property of the peculiar ZrW$_2$O$_8$ topology---is such that these fluctuations are described by a family of phonons necessarily spread throughout the BZ, with individual spaghetti distortions described by the population of a family of NTE phonons. Consequently a `spaghetti dynamics' mechanism for NTE in ZrW$_2$O$_8$ finally explains (i) why NTE modes are distributed throughout the BZ, (ii) why it is so difficult to interpret individual NTE phonons in isolation, and (iii) how the local polyhedral rotations/translations identified in previous studies \cite{Tucker_2005,Bridges_2014} actually propagate throughout the crystal lattice. Moreover, in determining this mechanism, we have established an unexpected link between the correlated structural disorder associated with NHE, on the one hand, and the property of NTE, on the other hand, that is of significant conceptual importance in the broader context of understanding disorder--property relationships in functional materials \cite{Overy_2016}.

\begin{acknowledgments}
The authors gratefully acknowledge financial support from the E.R.C. (Grant 279705), the E.P.S.R.C. (U.K.), the Leverhulme Trust (Grant No. RPG-2015-292), a CNRS/Oxford collaboration grant, and from the Marshall Foundation to P.M.M. This work benefitted from access to HPC platforms provided by a GENCI grant (A0030807069). ALG thanks J.~S.~O.~Evans (Durham), J.~P.~Attfield (Edinburgh), B. Slater (UCL), and T. Ledsam (Oxford) for valuable discussions.

\end{acknowledgments}

\end{document}